# Development of Glass Resistive Plate Chambers for INO

Satyanarayana Bheesette (for the INO collaboration)
*Tata Institute of Fundamental Research, Mumbai, 400005, India*

The India-based Neutrino Observatory (INO) collaboration is planning to build a massive 50kton magnetised Iron Calorimeter (ICAL) detector, to study atmospheric neutrinos and to make precision measurements of the parameters related to neutrino oscillations. Glass Resistive Plate Chambers (RPCs) of about 2m × 2m in size are going to be used as active elements for the ICAL detector. We have fabricated a large number of glass RPC prototypes of 1m × 1m in size and have studied their performance and long term stability. In the process, we have developed and produced a number of materials and components required for fabrication of RPCs. We have also designed and optimised a number of fabrication and quality control procedures for assembling the gas gaps. In this paper we will review our activities towards development of glass RPCs for the INO ICAL detector and will present results of the characterisation studies of the RPCs.

## 1. INTRODUCTION

ICAL is a 50kton magnetised iron tracking calorimeter, comprising of about 140 layers of low carbon 60 mm thick iron plates. Good tracking, energy and time resolutions as well as charge identification of the detecting particles are the essential capabilities of this detector. Sandwiched between these layers are glass RPCs, which are used as the active detector elements. About 27,000 RPCs of dimensions 2m × 2m will be deployed in this detector. Lateral dimensions of this cubical geometry detector are 48m × 16m × 12m. The geometry and structure of ICAL magnet which is used to produce a strong field of about 1.5Tesla, is largely fixed by the principle of ICAL detector. Its purpose is two fold; providing target nucleons for neutrino interactions and also a medium in which secondary charged particles can be separated on the basis of their magnetic rigidity [1][2].

## 2. OVERVIEW OF OUR EARLIER WORK

We have started our detector R&D work by fabricating several dozen glass RPCs of dimensions 30cm × 30cm. A gas mixing unit capable of mixing four individual gases and control the mixed gas flow through the detector chambers has been designed and developed. We have studied the basic operating characteristics of the chambers, such as RPC pulse profiles, voltage-current relationship, individual counting rates of the RPC. We have established a reliable way of monitoring the stability of the chamber based on its noise rates. We have obtained chamber plateau efficiencies of over 90% as well as made detailed measurements on the charge-time linearity, time response as well as the time resolution of the prototype RPCs.

We have then studied long-term stability of the glass RPC prototypes. These chambers were operated in avalanche gas mode, using a gas mixture of Freon (R134a) and Isobutane in the proportion 95.5:4.5 by volume and chamber operating voltage of 9.3KV. External amplification to the pickup signals was provided by preamplifiers. A comprehensive monitoring system for periodically recording the RPC high voltage currents as well as the ambient parameters such as temperature, pressure and relative humidity – both inside and outside the laboratory has been designed and implemented. Using this information along with the RPC test data, several correlations between the ambient parameters and the RPC operating characteristics could be established. The performance of these chambers, characterised by their efficiency, leakage currents and noise rates, had not changed over a period of three and a half years, thus establishing their long-term stability.

Finally, we had fabricated ten RPCs of dimensions 30cm × 30cm and mounted them as a stack such that the signal pickup strips of all the chambers were well aligned geometrically. Using even this stack of modest volume, we could record very interesting cosmic ray muon induced tracks. Muons arriving at different angles could be captured simply by relocating the telescope window. This has demonstrated that indeed these prototype chambers are capable of effectively tracking cosmic ray muons. The information recorded in these tests was also be used extract other parameters of interest, such as efficiency, noise rate and timing of individual RPCs and their long term stability.

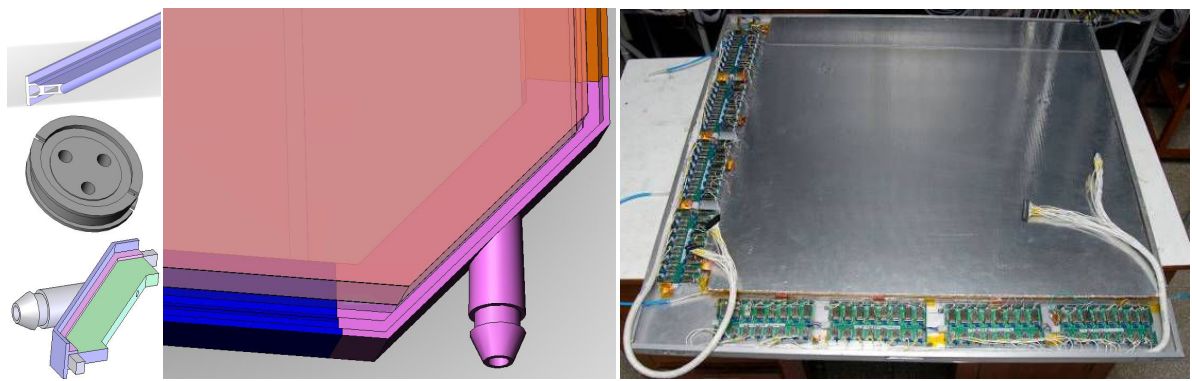

Figure 1: Polycarbonate based spacer, button and gas nozzle (left), conceptual assembly drawing of an RPC (middle), and fully assembled RPC of dimensions 1m × 1m (right)

## 3. DEVELOPMENT OF RPC MATERIALS AND ASSEMBLY PROCEDURES

RPC fabrication involves deploying a large number of materials as well as many assembly procedures. So, production of high performance and reliable chambers involves choosing the right type and quality of materials as well as optimisation of various assembly and quality control procedures involved in the fabrication. Materials such as glass used for electrodes, individual gases used for mixing and flowing the gas mixtures for the operation of the chambers, spacers, buttons, gas nozzles (see Figure 1) etc. which are needed for the assembling the chamber, resistive coat on the electrodes, epoxies used for gluing together different types of materials, pickup panels used for external signal pickup from the chambers, polyester films used for insulating the pickup panels from the resistive coated electrodes, to name some. We have studied a number of different type and quality of these materials and optimised most of these items.

We have also designed a developed a large number of assembly and quality control procedures and invented a number of useful jigs that are extremely useful in production of good quality detectors. Coating of semi-resistivity paint on the electrodes, assembling and gluing of chambers, leak testing of the finished chambers are some of the important assembly procedures. We have worked closely with various R&D institutions as well as many industrial houses in developing these materials and designing and developing the assembly procedures.

The RPCs are extremely disproportionate and heavy detector modules and hence pose serious problems in terms of mechanical rigidity and difficulties in the packing, transportation and installation of the modules in the detector. In addition, in spite of gluing the glass sheets together with a matrix of buttons throughout the area and using spacers on the four edges, the chamber tends to bulge outwards when the gas is flown through the chamber. These considerations call for a suitable and light weight casing for the chamber.

Plastic honeycomb panels laminated on one side by aluminum sheet and the other side by copper sheet was developed by us with help of an industry. Pickup strips of 30mm width are realised by machining the copper sheet. The honeycomb panels were found to be an excellent solution to the mechanical support to the RPC, since it offered better rigidity to the detector with much lesser weight than that of aluminum.

Resistive coating of the outer surfaces of the electrodes plays very crucial role in the operation of the RPC detector. We have collaborated with a local paint industry to develop a suitable paint as well as its application methods, which will be more adaptable for large scale production of the RPCs. We have also started designing paint automation techniques and developed prototype robotic machines for the purpose with help of local industry. We are also exploring the alternate and cost effective technique of coating the glass surface using screen printing method. This method can also be used to coat the paint on a PET film which can be stuck on the glass electrode surfaces [3].

### 4. CONSTRUCTION OF ICAL PROTOTYPE DETECTOR

The structure of the prototype detector is built in the form of a 13 layer sandwich of 50mm thick low carbon iron plates and 12 glass RPCs of 1m × 1m in area. The overall design of prototype magnet was kept as close to the conceived design of the full scale INO magnet. As the prototype magnet detects muons, it serves as the medium in which secondary charged particles can be separated on the basis of their magnetic rigidity. The detector is magnetised to 1.5Tesla, which enables momentum measurement of 1-10Gev muons produced by $\nu_\mu$ interactions within detector. Four sets of copper coils of 5 turns each, which were made from electrolytic copper conductor tubing having a central bore for flowing low conductivity water, are used for this purpose (see Figure 2).

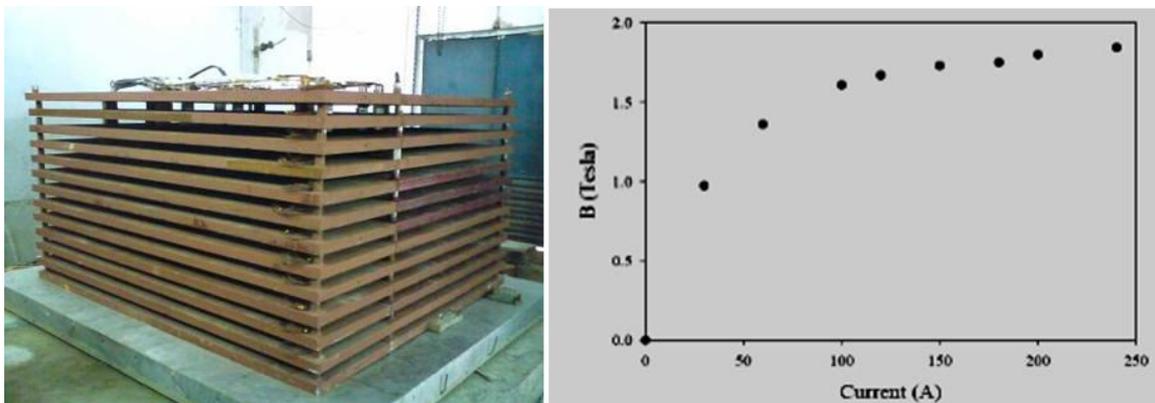

Figure 2: Prototype detector magnet in the assembled state (left) and its B-H curve (right)

A sophisticated gas mixing and distribution system, which works on four different input gases has been designed and fabricated. It features molecular sieve based filter columns on the input gas lines, Nippon Tylan made model FC-760 Mass Flow Controllers to precisely mix the gases to required proportion, Parker made fine filters and on-line moister readout on the mixed gas manifold. Mixed gas flow into 16 pneumatically controlled output channels is controlled by 0.3mm diameter stainless steel capillaries. Each of the output gas channels is equipped with a pair of bubblers, one on the detector input side for protecting the RPC in case of a gas channel block and the other on the output to isolate the chamber from the atmosphere. The gas system also features a facility to add controlled moister into the gas mixture, which is useful while working with the bakelite RPCs. The entire operation of control and monitoring of gas system is controlled by a PC using a dedicated hardware interface. All the input gas channels have been precisely calibrated over their entire dynamic range by water displacement and other methods [4]. The RPCs are operated in the avalanche mode using a gas mixture of R134:Isobutane:$SF_6$ in the proportion 95.5:5.0:0.5.

## 5. ELECTRONICS AND DATA ACQUISITION SYSTEMS

Role of the data acquisition system then is to generate the trigger, based on the hit pattern of the RPC pickup strips and to record strip hit patterns as well as timing of individual RPCs with reference to the trigger. Monitoring the stability of the detector as well as laboratory ambient parameters such as temperature, relative humidity and barometric pressure, are the other important tasks of the on-line data acquisition system.

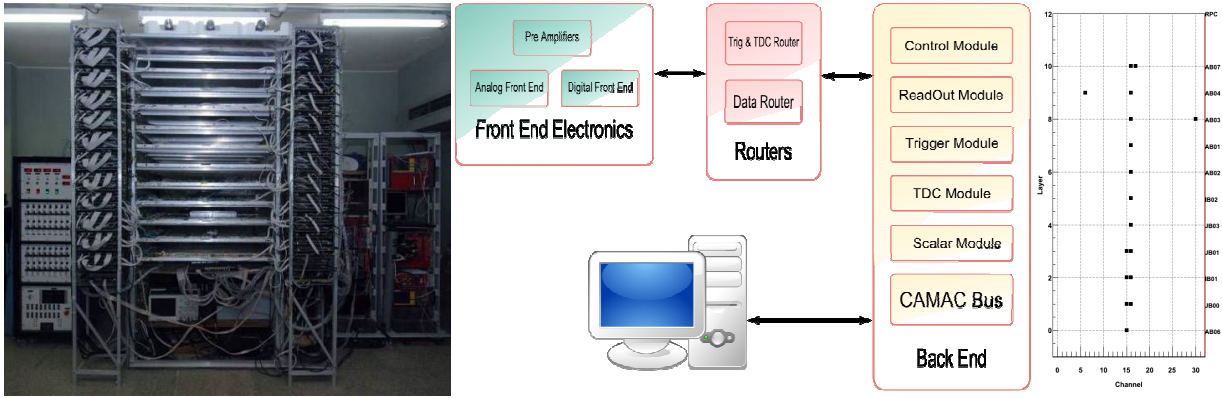

Figure 3: Prototype detector stack (left), schematic of its DAQ system (middle) and a muon track in the stack (right)

The signal readout chain essentially consists of a front-end fast high gain HMC based preamplifier and low level threshold discriminator followed by the digital front-end. The digital front-end built around a couple of CPLD chips handles important tasks of latching the strip hit pattern on master trigger as well as serially transferring the data to back-end. It is also here that the pre-trigger signals are generated, which are used by the back-end trigger module built using combinatorial circuits and produces master trigger. Finally, the digital front-end system also handles the entire signal multiplexing required both during the event data acquisition as well as during strip signal rate monitoring. The timing data is acquired using the commercial TDC modules. The data acquisition is done using a CAMAC backend, employing many custom built modules, such as control and readout modules. The multiplexed signals from the front-end are sent to the back-end through appropriate router modules. Monitoring of strip signal rates is done as a cyclic background job using the scaler module in the back-end. Monitoring of ambient laboratory parameters such as temperature, relative humidity and barometric pressure along with important operating parameters of the RPCs such as high voltage and current is also routinely done and made available on-line on web. Shown in Figure 3 are the prototype detector stack of 12 RPCs, schematic of its data acquisition system and typical muon track in the stack.

## 6. RESULTS

The prototype detector stack is in un-interrupted operation now for about a year and the data is being acquired using the on-line system described above. The recorded data is stored in a customised format by the on-line software and is analysed off-line in detail using sophisticated ROOT based analysis software. Some of the aspects that are analysed on day to day basis are the RPC efficiencies for cosmic ray muons, absolute and relative timing resolutions as well as the stability of RPCs based on the monitoring data of the individual strip rates. Timing resolution plots for four individual RPCs is shown in Figure 4. These RPCs show timing resolutions of about 2nSec with reference to scintillator paddle based trigger signal. Shown in Figure 4 are also the strip rate monitor profiles of an RPC. As can be seen from the plots, the noise rates are very stable over the period of monitoring inferring the stability of operation of the RPCs under test.

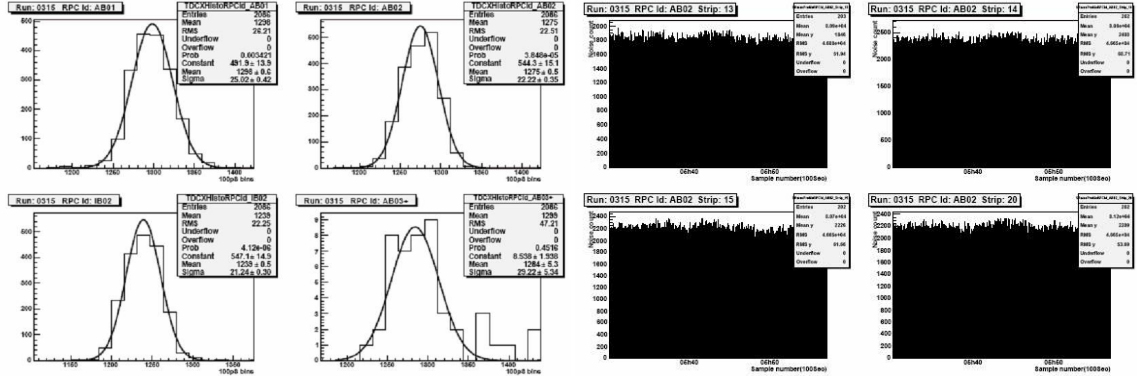

Figure 4: Time resolution (left) and noise rate monitor profile (right) plots of RPCs

## 7. CONCLUSIONS AND OUTLOOK

Large area RPCs of dimensions 1m × 1m required for the ICAL prototype detector were successfully developed, built and characterised. They were operated stably in avalanche mode for long period of time without any signs of aging and other problems. Electronics, trigger, data acquisition and monitoring system hardware required to operate the detector has been designed and developed indigenously and commissioned. Necessary data analysis tools have also been developed. The prototype detector magnet has been designed, fabricated, installed and was found to produce the designed field. Final integration of the magnet, the active detector elements and the electronics systems is in progress and we expect to start acquiring the data from the prototype detector soon.

## Acknowledgments

The author would like to thank all his INO collaboration colleagues for their valuable contributions towards the work reported in this paper.